\documentclass[12pt,a4paper,final]{iopart}
\usepackage{iopams}

\expandafter\let\csname equation*\endcsname\relax
\expandafter\let\csname endequation*\endcsname\relax

\usepackage{amsmath}
\usepackage{amsfonts}
\usepackage{amsmath}
\usepackage{amssymb}
\usepackage{amsthm}
\usepackage[utf8]{inputenc}
\usepackage{bm}
\usepackage{mathrsfs}
\usepackage{graphicx}
\usepackage{comment}

\usepackage[breaklinks=true,colorlinks=true,linkcolor=blue,urlcolor=blue,citecolor=blue]{hyperref}

\newcommand{\defeq}{\mathrel{\mathop:}=}

\renewcommand{\H}{\mathcal{H}}
\newcommand{\D}{\mathcal{D}}
\newcommand{\nanika}{{\scriptstyle \bullet}}

\begin{document}

\title{Configurational density of states of power-law potentials and the virial theorem in steady states}

\author[cor1]{Sergio Davis$^{1,2}$}
\address{$^1$Research Center on the Intersection in Plasma Physics, Matter and Complexity (P$^2$mc), Comisión Chilena de Energía Nuclear, Casilla 188-D, Santiago, Chile}
\address{$^2$Departamento de Física y Astronomía, Facultad de Ciencias Exactas, Universidad Andres Bello. Sazié 2212, piso 7, 8370136, Santiago, Chile.}
\ead{sergio.davis@cchen.cl}

\begin{abstract}
In this brief note, the configurational density of states of a system of particles interacting via power-law pair potentials is computed exactly. The result is consistent with a constant microcanonical 
heat capacity. The well-known form of the virial theorem for this class of systems is recovered using only the obtained configurational density of states, and shown to be valid beyond the canonical and 
microcanonical ensembles, in general steady states.
\end{abstract}

\section{Introduction}

The virial theorem~\cite{Greiner2012} is a well-known result of equilibrium statistical mechanics. For a system of $N$ classical particles with Hamiltonian
\begin{equation}
\H = K(\bm{p}_1, \ldots, \bm{p}_N) + \Phi(\bm{r}_1, \ldots, \bm{r}_N),
\end{equation}
with
\begin{equation}
\label{eq:kin}
K(\bm{p}_1, \ldots, \bm{p}_N) \defeq \sum_{i=1}^N \frac{\bm{p}_i^2}{2 m_i}
\end{equation}
the total kinetic energy of the system and $\Phi$ the potential energy including arbitrary interactions, one of the many forms of the virial theorem is the equality
\begin{equation}
\label{eq:virial}
\big<K\big>_0 = \frac{1}{2}\left<\sum_{i=1}^N \bm{r}_i\cdot \frac{\partial \Phi}{\partial \bm{r}_i}\right>_0
\end{equation}
where the notation $\big<\nanika\big>_0$ denotes either canonical or microcanonical equilibrium expectation values. If the interaction potential $\Phi$ is the sum of pair interactions, that is, if
\begin{equation}
\label{eq:phi_pair}
\Phi(\bm{r}_1, \ldots, \bm{r}_N) = \frac{1}{2}\sum_{i=1}^N \sum_{j \neq i} \varphi\big(\big|\bm{r}_j-\bm{r}_i\big|\big),
\end{equation}
where $\varphi$ is a function only of the interatomic distance, and is of the form
\begin{equation}
\label{eq:pair_power}
\varphi(r) = \varphi_0\,r^\gamma
\end{equation}
with $\varphi_0$ and $\gamma$ constants, then it is possible to show that
\begin{equation}
\label{eq:virial_pair}
\big<K\big>_0 = \frac{\gamma}{2}\big<\Phi\big>_0.
\end{equation}

\section{Configurational density of states}

Let $\D(\phi)$ denote the configurational density of states (CDOS) associated to the interaction potential $\Phi$, defined by
\begin{equation}
\label{eq:cdos}
\D(\phi) \defeq \int d\bm{r}_1\ldots d\bm{r}_N\;\delta\big(\Phi(\bm{r}_1, \ldots, \bm{r}_N)-\phi\big).
\end{equation}

\noindent
Replacing the potential energy $\Phi$ with \eqref{eq:phi_pair} and \eqref{eq:pair_power}, we can write
\begin{equation}
\D(\phi) = \int d\bm{r}_1\ldots d\bm{r}_N\;\delta\left(\sum_{i=1}^N \sum_{j\neq i}\varphi_0\big|\bm{r}_j-\bm{r}_i\big|^\gamma - \phi\right),
\end{equation}
and by extracting $\phi$ out of the Dirac delta we have
\begin{equation}
\D(\phi) = \frac{1}{\phi}\int d\bm{r}_1\ldots d\bm{r}_N\;\delta\left(1 - \sum_{i=1}^N \sum_{j \neq i}\frac{\varphi_0}{\phi}\big|\bm{r}_j-\bm{r}_i\big|^{\gamma}\right),
\end{equation}
which suggest the change of variables
\begin{equation}
\bm{u}_i \defeq \left(\frac{\varphi_0}{\phi}\right)^{\frac{1}{\gamma}}\bm{r}_i
\end{equation}
in order to extract the full dependence of $\D(\phi)$ on $\phi$. In fact, writing the integral in terms of $\bm{u}_1, \ldots, \bm{u}_N$ we have
\begin{equation}
\D(\phi) = \frac{1}{\varphi_0}\left(\frac{\phi}{\varphi_0}\right)^{\frac{3N}{\gamma}-1}\;\left[\int d\bm{u}_1\ldots d\bm{u}_N\;\delta\left(1-\sum_{i=1}^N \sum_{j \neq i}\big|\bm{u}_j-\bm{u}_i\big|^{\gamma}\right)\right]
\end{equation}

\begin{equation}
\label{eq:cdos_pair}
\D(\phi) = \frac{1}{\varphi_0}D_N(\gamma)\,\left(\frac{\phi}{\varphi_0}\right)^{\frac{3N}{\gamma}-1},
\end{equation}
where $D_N(\gamma)$ is a constant which does not affect the thermodynamic properties. Immediately we see that, for $\gamma = 2$, this result agrees with the CDOS of a system of $N$ harmonic oscillators.

\section{The virial theorem}

The virial theorem, as in \eqref{eq:virial_pair}, will be derived using \eqref{eq:cdos_pair} for an arbitrary steady state of the form
\begin{equation}
\label{eq:rho}
P(\bm{R}, \bm{P}|S) = \rho\big(K(\bm{P}) + \Phi(\bm{R}); S),
\end{equation}
where $\rho$ is the \emph{ensemble function}, $\bm{R} \defeq (\bm{r}_1, \ldots, \bm{r}_N)$, $\bm{P} \defeq (\bm{p}_1, \ldots, \bm{p}_N)$ and $K(\bm P)$ is given by \eqref{eq:kin}. The joint distribution 
of kinetic and potential energy in the state $S$ is then given by
\begin{equation}
\begin{split}
P(k, \phi|S) & = \Big<\delta(K(\bm P)-k)\,\delta(\Phi(\bm R)-\phi)\Big>_S \\
& = \rho(k+\phi; S)\left[\int d\bm{P}\delta\big(K(\bm P)-k\big)\right] \left[\int d\bm{R}\delta\big(\Phi(\bm R)-\phi\big)\right],
\end{split}
\end{equation}
that is, we have
\begin{equation}
P(k, \phi|S) = \rho\big(k+\phi; S\big)\Omega_K(k; N)\D(\phi)
\end{equation}
with $\Omega_K(k; N)$ the kinetic density of states, given by
\begin{equation}
\Omega_K(k) \defeq \int d\bm{p}_1\ldots d\bm{p}_N\,\delta\left(\sum_{i=1}^N \frac{\bm{p}_i^2}{2 m_i} - k\right) = W_N\,k^{\frac{3N}{2}-1}
\end{equation}
with $W_N$ a constant whose value is
\begin{equation}
W_N = \frac{1}{\Gamma\left(\frac{3N}{2}\right)}\prod_{i=1}^N (2\pi m_i)^{\frac{3}{2}}.
\end{equation}

Applying the conjugate variables theorem~\cite{Davis2012, Davis2016} to the joint distribution $P(k, \phi|S)$ we have
\begin{align}
\label{eq:CVT_k}
\left<\frac{\partial \omega}{\partial K}\right>_S & = \left<\omega\left(\beta_F - \frac{3N-2}{2K}\right)\right>_S, \\
\label{eq:CVT_phi}
\left<\frac{\partial \omega}{\partial \Phi}\right>_S & = \left<\omega\left(\beta_F - \frac{3N-\gamma}{\gamma \Phi}\right)\right>_S
\end{align}
for $\omega = \omega(K, \Phi)$ an arbitrary differentiable function of $K$ and $\Phi$, and where we have defined the \emph{fundamental inverse temperature}
\begin{equation}
\beta_F(E; S) \defeq -\frac{\partial}{\partial E}\ln \rho(E; S).
\end{equation}

\noindent
Using \eqref{eq:CVT_k} with $\omega(\phi, k) = \phi\cdot k$, we have
\begin{equation}
\label{eq:vir1}
\big<\beta_F\,\Phi\,K\big>_S = \frac{3N}{2}\big<\Phi\big>_S
\end{equation}
while \eqref{eq:CVT_phi} with $\omega(\phi, k) = \phi\cdot k$ yields
\begin{equation}
\label{eq:vir2}
\big<\beta_F\,\Phi\,K\big>_S = \frac{3N}{\gamma}\big<K\big>_S,
\end{equation}
thus by equating the right-hand sides of \eqref{eq:vir1} and \eqref{eq:vir2} we prove \eqref{eq:virial_pair} in the more general form
\begin{equation}
\big<K\big>_S = \frac{\gamma}{2}\big<\Phi\big>_S,
\end{equation}
valid for any steady state where \eqref{eq:rho} holds. The total density of states is given by
\begin{equation}
\begin{split}
\Omega(E) & = \int d\bm{R}d\bm{P}\,\delta\big(K(\bm P) + \Phi(\bm R) - E\big) \\
& = \int_0^\infty d\phi \int_0^\infty dk\,\Omega_K(k)\D(\phi)\delta(k+\phi-E) \\
& = \int_0^\infty d\phi\,\Omega_K(E-\phi)\D(\phi) \\
& = \frac{W_N D_N(\gamma)}{\varphi_0}\int_0^E d\phi\,(E-\phi)^{\frac{3N}{2}-1} \phi^{\frac{3N}{\gamma}-1} = \frac{W_N\,D_N(\gamma)}{\varphi_0}\,
B\Big(\frac{3N}{2}, \frac{3N}{\gamma}\Big)E^{\frac{3N}{2}+\frac{3N}{\gamma}-1}
\end{split}
\end{equation}
with $B(a, b)$ the beta function. The microcanonical inverse temperature is then
\begin{equation}
\frac{1}{k_B T(E)} = \frac{\partial}{\partial E}\ln \Omega(E) = \frac{1}{E}\left(\frac{3N}{2}+\frac{3N}{\gamma}-1\right),
\end{equation}
so the system has constant microcanonical heat capacity
\begin{equation}
C_E = \left(\frac{3N}{2} + \frac{3N}{\gamma} - 1\right)k_B.
\end{equation}

The stability of the entire system in the canonical ensemble requires $C_E > 0$. This is possible for any positive $\gamma$, as well as for negative $\gamma$ if
\begin{equation}
\gamma < -\frac{6N}{3N-2},
\end{equation}
thus ruling out the gravitational potential which has $\gamma = -1$.

\section*{Acknowledgments}

Funding from ANID FONDECYT 1220651 is gratefully acknowledged.

\section*{References}

\bibliography{virial}
\bibliographystyle{unsrt}

\end{document}